\documentclass{iau}
\usepackage{graphicx,natbib}
 \usepackage{subcaption}
 
\newcommand{\apj}{ApJ}           
\newcommand{\apjl}{ApJ}           
\newcommand{\mnras}{MNRAS}       

\newcommand{\aap}{A\&A}
\newcommand{\araa}{ARA\&A}
\newcommand{\aj}{AJ}

\def\msun{$M_\odot$}
\def\sigsfr{$\Sigma_{SFR}$\ }
\def\siggas{$\Sigma_{gas}$\ }
\def\siggasp{$\Sigma_{gas}$}

\title{On Schmidt's Conjecture and Star Formation Scaling Laws}

\author[Lada]{Charles J. Lada$^1$}

\affiliation{$^1$Harvard-Smithsonian Center for Astrophysics,\\60 Garden Street, Cambridge, MA 02138 USA  \\
email: {\tt clada@cfa.harvard.edu} \\
}

\pubyear{2014}
\volume{309}
\jname{Galaxies in 3D across the Universe}
\editors{B. L. Ziegler, F. Combes, H. Dannerbauer, M. Verdugo, eds.}

\begin{document}

\maketitle

\begin{abstract}
Ever since the pioneering work of Schmidt a half-century ago there has been great interest in finding an appropriate empirical relation that would directly link some property of interstellar gas with the process of star formation within it. Schmidt conjectured that this might take the form of a power-law relation between the rate of star formation (SFR) and the surface density of interstellar gas. However, recent observations  suggest that a linear scaling relation between the total SFR and the amount of dense gas within molecular clouds appears to be the underlying physical relation that most directly connects star formation with interstellar gas from scales of individual GMCs to those encompassing entire galaxies both near and far. Although Schmidt relations are found to exist within local GMCs, there is no Schmidt relation observed between GMCs. The implications of these results for interpreting and understanding the Kennicutt-Schmidt scaling law for galaxies are discussed.
\end{abstract}

\firstsection
\section{Introduction}

One of the fundamental challenges confronting modern day astrophysics is the determination of the physical processes that control the conversion of  interstellar gas into stars.  Understanding of this process is a key to deciphering both star formation and galaxy evolution across cosmic time. A little more  than half a century ago Schmidt (1959) conjectured that "It would seem probable that the rate of star formation depends on the gas density and...varies with a power-law
 of the gas density." Or as later parameterized by Kennicutt (1998): $\Sigma_{SFR} = \kappa \Sigma_{gas}^n$,  where $\Sigma_{SFR}$ and $\Sigma_{gas}$ are {\it surface} densities of the respective quantities.  Based on a number considerations, including the local distributions over $z$ of B stars and HI gas, Schmidt argued that such a scaling relation existed in the solar neighborhood and that n $\approx 2$. He also noted that "It is rather tempting to try to estimate the effects of star formation...in galaxies as a whole." However, it would take more than four decades before astronomers could adequately test Schmidt's idea in galaxies. It turns out that it was not so easy to determine accurate star formation rates for whole galaxies when what one actually measures is a flux of the galaxy at one or perhaps a number of different wavelengths. The solution to this problem required the development of population synthesis modeling in order to reasonably predict star formation rates from observed fluxes. It wasn't until the late 1970s and early 1980s that technology and astrophysical knowledge of such issues as stellar evolution, stellar spectra, stellar IMFs, dust, etc. reached the necessary level to enable meaningful determinations of star formation rates in galaxies using population synthesis calculations.   Moreover, it wasn't until the late 1970s that the molecular gas content in galaxies could be measured using CO and thus provide a complete census of both HI and H$_2$ gas surface densities for comparison with the galaxy-wide SFRs. By 1989, nearly 30 years after Schmidt first published his conjecture, observations of blue stars, HII regions and gas surface densities confirmed the existence of Schmidt-like relations for disk galaxies but with  a large dispersion in the value of the power-law index, $0 \leq n \leq 4$ (Kennicutt 1989 and references therein). Another decade later it became possible for Kennicutt  (1998) to put together existing observations and models and provide a more robust test of Schmidt's conjecture for whole  galaxies. In his classic paper he showed that a power-law scaling relation (now known as the Kennicutt-Schmidt law) exists between star forming galaxies and is furthermore characterized by an index, n = 1.6. Twenty years later observational capabilities improved to the point where it was possible to investigate the Kennicutt-Schmidt law {\it within} individual galaxies on kpc spatial scales. In an influential paper, Bigiel et al. (2008) produced a Kennicutt-Schmidt diagram that combined resolved observations of local disk galaxies with more global observations of local starburst galaxies. They found a more complex scaling between \sigsfr and \siggas that could not simply be  described by a single power-law (see Figure \ref{fig:ks_betal}). Given this development it is interesting to consider what the Schmidt relation would be if we could make measurements on smaller, sub-kpc scales, indeed, even on the spatial scales of giant molecular clouds (GMCs) themselves.
 \begin{figure}
\centering
\includegraphics[width=0.55\columnwidth]{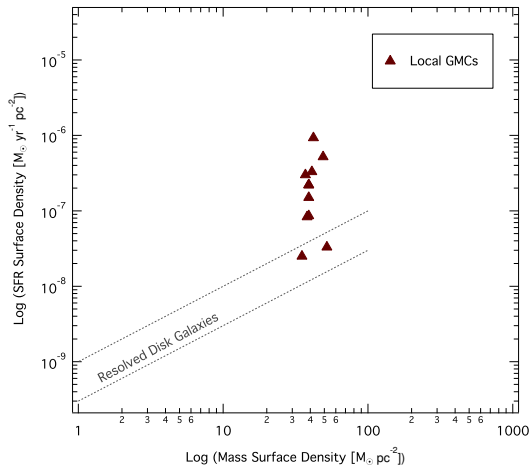} 
\caption{The Kennicutt-Schmidt diagram for nearby GMCs.  There is no Schmidt scaling relation between local GMCs.   The parallel dotted lines represent the approximate locus of (resolved) disk galaxies which can be described by a linear scaling law in the Kennicutt-Schmidt plane (Schruba et al. 2011). GMC data from Lada et al. (2010). See text.
}
\label{fig:fig1}
\end{figure}
 \section{The Schmidt Law on GMC scales}
 
 Over the last decade wide-field infrared surveys from the ground and space have made it possible to obtain both accurate masses of local molecular clouds (from infrared extinction measurements) and nearly complete inventories of the populations of young stellar objects (YSOs) contained within these clouds (from both ground and space-based infrared, optical and x-ray observations). Furthermore, ages and masses of these YSOs can be derived from their positions on the HR diagram. And, for clouds within 0.5 kpc of the sun, the stellar IMF is sampled over its entire range, from OB stars down to the hydrogen burning limit. Using these data  fairly accurate determinations of cloud sizes, masses and star formation rates (SFRs) are possible. Combining these measurements, the individual GMCs can all be placed on the Kennicutt-Schmidt (KS) diagram. Figure 1 shows the result of this exercise for a nearly complete sample of molecular clouds within 0.5 kpc of the sun (Lada et al. 2013). Clearly, there is {\it no} Schmidt scaling relation between local GMCs.  This result is perhaps surprising at first look, but it is in fact a consequence of the well know scaling relation between mass and radius (M$_{\rm GMC}$ $\propto$ R$^2$) of Milky Way (MW)  GMCs, discovered over three decades ago by Larson (1981).  Yet, in the context of the observed KS relation for disk galaxies (e.g., Schruba et al., 2011), the lack of a Schmidt relation for GMCs is nonetheless intriguing. 
 \begin{figure}[t]
\centering
\begin{subfigure}{.4\textwidth}
\centering
\includegraphics[width=\textwidth]{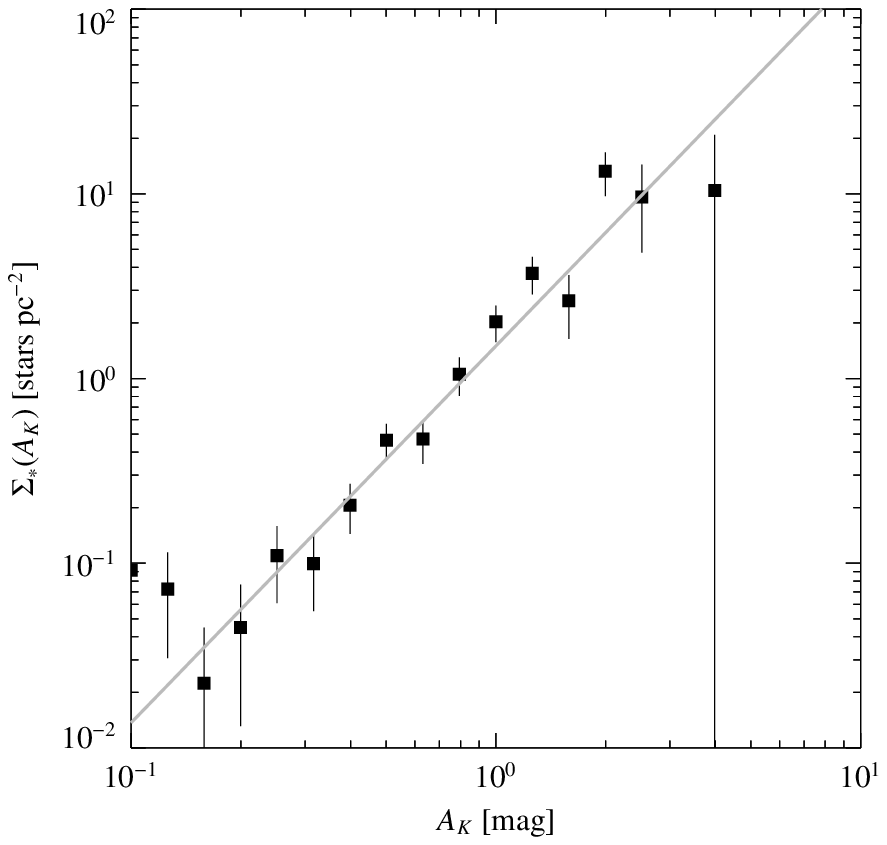} 
\end{subfigure}
\begin{subfigure}{.4\textwidth}
\centering
\includegraphics[width=\textwidth]{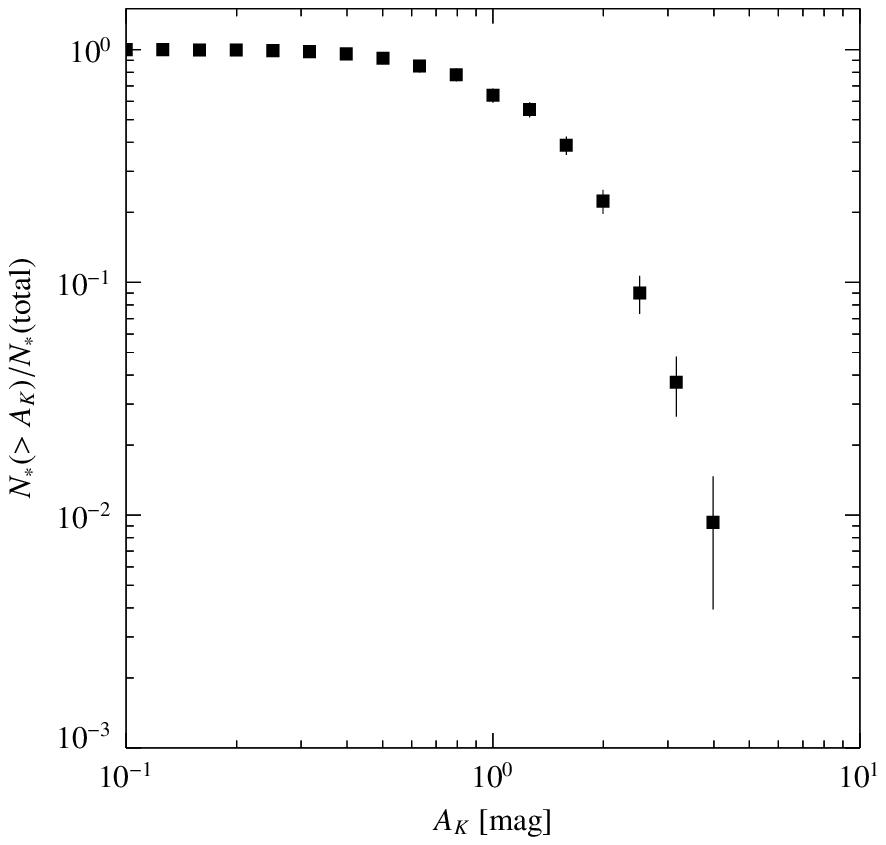} 
\end{subfigure}
\caption{The left panel shows that a Schmidt scaling relation does exist within a single GMC, e.g., the Orion A cloud. Here $\Sigma_*$ $\propto$ $A{_V}^{2.0}$, where $A_V$ is the dust surface density, a proxy for $\Sigma_{H_2}$. The right panel shows the cumulative protostellar fraction (or fractional yield of star formation) in the cloud as a function of $A_V$. From Lada et al. (2013).
}
\label{fig:fig2}
\end{figure}
 
 \section{The Schmidt Law on sub-GMC scales}
 
 Because local GMCs are so well resolved by observations, it is interesting to next consider whether  a Schmidt scaling relation exists {\it within} clouds (i.e., on sub-GMC or pc scales). Indeed, a series of recent observational studies have demonstrated that a Schmidt scaling relation does exist within local clouds (Heidermann et al. 2010, Gutermuth et al. 2011, Lombardi et al. 2013, Lada et al. 2013, Evans et al. 2014). As an example, Figure \ref{fig:fig2} (left panel) shows the Schmidt relation for the Orion A GMC derived by Lada et al. (2013) from highly resolved observations of the protostellar surface density,  $\Sigma_*$, and extinction, $A_V$ (i.e., dust surface density). These two quantities are directly proportional to \sigsfr and $\Sigma_{H_2}$, respectively. Here a fit to the data gives $\Sigma_*$ $\propto$ $A{_V}^{2.0}$. The unambiguous presence of a Schmidt scaling relation on sub-GMC scales sharply contrasts with the unambiguous lack of such a relation on GMC scales. How can one reconcile these different results derived using the same set of observations? 
 
 Protostars are the youngest, least evolved YSOs and can be thought of as the instantaneous product of the star formation process. The right panel of Figure \ref{fig:fig2} shows the (cumulative) fractional number of protostars plotted as a function of extinction in the Orion GMC.  Despite the fact that $\Sigma_*$ rises so steeply with extinction, the actual number of protostars produced by the cloud sharply decreases with extinction.  This is equivalent to the statement that although \sigsfr increases non-linearly with extinction, the overall SFR decreases sharply with extinction. This seemingly paradoxical situation is resolved when one considers that, 1) the total star formation rate is given by: $ {\rm SFR} = \int \Sigma_{SFR} dS$,  the integral of the surface density over the surface area, S, and 2) the surface area, $S$($A_K$), of a cloud sharply decreases with increasing extinction (Lada et al. 2013).  Indeed, it is the coupling of these two functions,  \sigsfr($A_K$) and $S(A_K)$ that sets the global star formation rate in a cloud. Clearly, cloud structure plays a fundamental role in determination of its star formation rate. In this context the Schmidt scaling relation can be considered a differential scaling relation and is, by itself, not a reliable predictor of the level of star formation in a cloud. Two clouds of the same mass and size and characterized by the same Schmidt scaling relation will not necessarily produce similar levels of star formation. Since most clouds seem to be characterized by Schmidt relations with similar power-law indices (n $\approx$ 2; Lada et al. 2013) observed differences in their SFRs (Figure 1) are primarily the result of differences in their surface area distribution functions, $S(A_K)$. 
 
 \begin{figure}
\centering
\includegraphics[width=0.5\columnwidth]{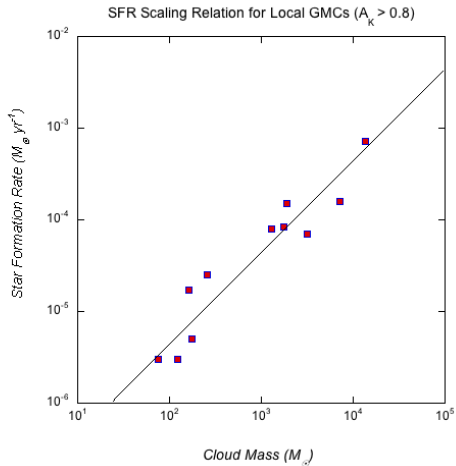} 
\caption{The scaling relation between (integrated) SFR and high extinction, presumably dense, gas mass for local GMCs.  Adapted from Lada et al. (2010). See text.
}
\label{fig:fig3}
\end{figure}

\section{Dense Gas and Integral Scaling Laws}

Close inspection of the right panel in Figure 2 shows that although the fractional number of protostars sharply decreases with extinction, 80\% of all protostars are found at extinctions in excess of $A_K$ $\approx$ 0.8 magnitudes. More recent observations using higher angular resolution Herschel observations (Lombardi et al. 2014) to map dust column density show that 90\% of all protostars in Orion lie above this extinction level.  This suggests that star formation is intimately related to high opacity and presumably high (volume) density material. Indeed, for local clouds there is a relatively tight scaling relation between the mass of high opacity gas and the global SFR (Lada et al. 2010). This (integral) scaling relation is shown in Figure 3, where the global SFR is plotted against the mass of the cloud contained above an extinction threshold of $A_K$ = 0.8 magnitudes. Unlike typical Schmidt relations, this scaling follows a linear relation, i.e., n = 1. This has been interpreted to mean that the SFR in a cloud is simply controlled by the amount of high extinction (and presumably dense) material contained within it. A similar linear scaling between the total gaseous mass in a cloud and its global SFR has also been found to exist (Lada et al. 2012), but it is characterized by a significantly higher dispersion (an order of magnitude in SFR). The increased dispersion in the latter relation compared to that of the dense gas scaling relation has been shown to be entirely due to the differing dense gas fractions in the clouds. 

\begin{figure}
\centering
\includegraphics[width=0.6\columnwidth]{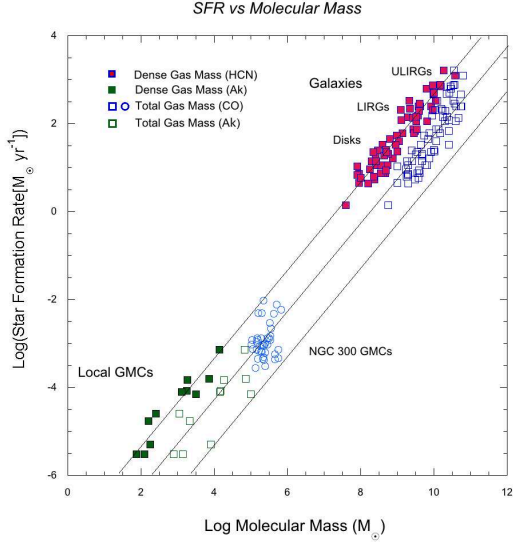} 
\caption{The integral scaling relation (SFR vs M($H_2$)) for GMCs and galaxies  for both dense gas (filled symbols) and total gas (open symbols) masses.  The dotted parallel lines represent lines of constant depletion times corresponding to 22 Myr, 220 Myr and 2.2 Gyr, from top to bottom, respectively, or alternatively, constant dense gas fractions of 100\%, 10\% and 1\%, respectively. See text.
}
\label{fig:fig4}
\end{figure}

The linear scaling between dense gas mass and the SFR in local clouds is reminiscent of a similar global relation found for galaxies by Gao and Solomon (2004) who showed that FIR luminosities of galaxies (including disk galaxies and nuclear starbursts) were linearly correlated with the luminosities of HCN molecular-line emission, a tracer of dense (n$_{H_2}$ $>$ 10$^4$ cm$^{-3}$) molecular gas. Subsequent observations by Wu et al. (2005) comparing FIR and HCN luminosities of massive GMCs in the Milky Way also showed a linear correlation between the two quantities and this relation was found to extrapolate smoothly to that found by Gao and Solomon for galaxies, spanning a range in scale of over nine orders of magnitude.  This can be seen in Figure \ref{fig:fig4} where the integral scaling relations (i.e., SFR vs M($H_2$)) for both dense gas and total mass measurements of MW GMCs (Lada et al 2010) and galaxies (Gao \& Solomon 2004) are presented on the same plot. Also shown are recent observations of GMCs in the nearby galaxy NGC 300, obtained on a $\sim$ 200 pc spatial scale (Faesi et al. 2014).   
Gas masses were obtained from extinction measurements for MW GMCs and from CO (total) and HCN (dense) gas measurements for the extragalactic objects. SFRs were determined from direct counting of YSOs in local GMCs and from population synthesis analyses for the extragalactic sources with appropriate calibrations (Chomiuk \& Povich 2011, Lada et al. 2012, Faesi et al. 2014). 

Qualitatively, the relationship between SFR scaling relations for dense and total gas masses found in local GMCs also appears  to characterize the observations of galaxies. This similarity suggests that we are observing a similar physical process in star forming environments across all spatial scales. In particular, the linear scaling between SFR and dense gas mass suggests that the rate of star formation is directly controlled by the amount of dense gas that can be assembled in a star forming region. 

The scaling relation between SFR and total cloud masses is also seemingly linear from GMCs to disk galaxies but appears to depart from this linearity at the highest SFRs, in the region occupied by active starbursts (i.e., ULIRGs). Lada et al. (2012) argued that this departure from a single linear relation was the result of the fact that the dense gas fractions in these starburst galaxies are systematically higher than those that characterize GMCs and disk galaxies.  They postulated the existence of a family of linear scaling laws between SFRs and total gas masses that are parameterized by the dense gas fraction in the star forming material. In other words, the star formation scaling law is always linear for systems with the same dense gas fraction. It is interesting to note that even galaxies at high z ($>$ 2-4) are characterized by similar scaling relations when FIR and CO luminosities  are compared (Genzel et al. 2010, Shapley 2011) further suggesting that the physical process of star formation in distant galaxies and even through much of cosmic time may be similar in some fundamental respect to that we are observing today in local GMCs.

 \begin{figure}
\centering
\includegraphics[width=0.6\columnwidth]{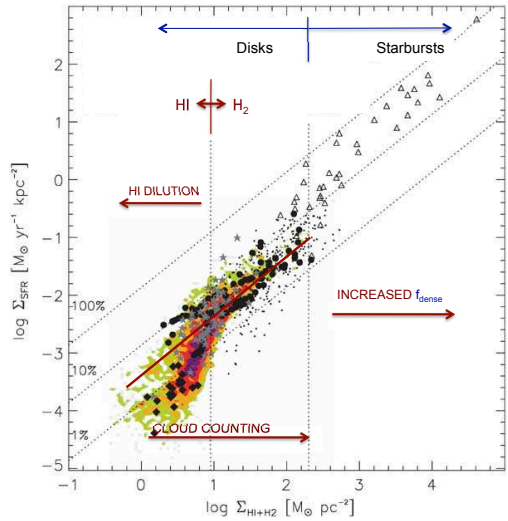} 
\caption{Interpreting the Kennicutt-Schmidt diagram for nearby disk and  starburst galaxies. Figure adapted from Bigiel et al. (2008). The solid brown line approximately represents the linear relation between $\Sigma_{\rm SFR}$ and $\Sigma_{H_2}$ from $\sim$ 1 \msun\  pc$^{-2} $ to $\sim$ 100 \msun\ pc$^{-2}$ found by Schruba et al. (2011). See text for discussion.
}
\label{fig:ks_betal}
\end{figure} 
 \section{The Nature of the Kennicutt-Schmidt Law for Galaxies}
 
Determinations of the KS relation for disk galaxies likely consist of measurements of star forming GMCs  and GMC complexes on spatial scales from 1 - 10 kpc. These measurements clearly display power-law scaling relations. How is this reconciled with the observations in Figure 1 that show no Schmidt relation for MW GMCs? Figure \ref{fig:ks_betal} primarily shows  resolved observations of disk galaxies in the KS plane (\sigsfr vs \siggas) from Bigiel et al. (2008). Unlike the situation for global measurements of such galaxies (Kennicutt 1998), a single power-law relation cannot adequately match the data. The relation is more complex and perhaps could be characterized by multiple (3) power-law relations. In order to interpret this diagram, we note that for \siggas $<$ 10 \msun\ pc$^{-2}$, the gas in the disks is predominately atomic (HI) in form, and above this threshold the gas is almost exclusively in molecular (H$_2$) in form. Star formation occurs exclusively in molecular gas, thus HI gas is inert to star formation. The combination of HI and H$_2$ gas in the measurement of \siggas tends to dilute the star formation signal in disks dominated by HI, (with \siggas $<$ 10 \msun\ pc$^{-2}$ ). This artificially steepens the slope between \sigsfr and \siggas.  The functional form of the scaling relation in this part of the diagram provides no insight into the actual physical process of star formation operating in the molecular gas.

For \siggas $>$ 10 \msun \ pc$^{-2}$, the scaling relation appears to be linear up to $\approx$ 100  \msun \ pc$^{-2}$.  Using an improved analysis of the data  in which the molecular observations of the galaxies were stacked to improve the sensitivity, Schruba et al. (2011) showed that if only CO observations are considered, this linear relation extends down to $\Sigma_{H_2}$ $\approx$ 1 \msun \ pc$^{-2}$. However, these observations are not consistent with the fact that GMCs are thought to be characterized by a constant column density of $\sim$ 40 - 50 \msun \ pc$^{-2}$\ (Figure 1 and Larson 1981). Moreover, a molecular cloud cannot exist if $\Sigma_{H_2}$ $<$ 16 \msun \ pc$^{-2}$, the threshold surface density required for self-shielding against UV photo-dissociation. Consequently, the vast majority of the (CO derived) measurements of $\Sigma_{H_2}$ likely represent beam diluted observations. Consider that in a 1 kpc square pixel a single GMC would be characterized by a filling factor, f $\sim$ 0.002. Consequently, the observed linear scaling relation between 1 -- 100 \msun \ pc$^{-2}$ likely represents a cloud-counting sequence which measures the surface density of GMCs rather than the surface density of gas within them.  As a result, the linear KS relation between 1 - 100 \msun \ pc$^{-2}$ reveals little about physics of the underlying  process of star formation within the molecular gas. 

In the portion of the KS diagram for which \siggas $>$ 100 \msun \ pc$^{-2}$ there appears to be a distinct break and jump to higher \sigsfr followed by a more or less linear scaling between \sigsfr and \siggasp. These observations represent global observations of primarily starburst galaxies. We interpret this shift as being a consequence of an increased dense gas fraction in these objects. Consider that in typical starburst nuclei, the mass of molecular gas can be similar to or larger than that contained in the entire disk of a normal star forming spiral galaxy (see Figure 4),  but  confined to a region only $\sim$ 1 kpc in size. The geometric decrease in surface area alone would be enough to significantly increase \sigsfr and \siggas compared to normal galaxies. In addition the mean volume density of the material must also be significantly enhanced which in turn would lead to increases in the global SFRs for the molecular gas in these systems compared to that in GMCs. Interestingly, observations suggest that the molecular gas is still beam diluted by observations on these scales, but this likely cannot account fully for the linear nature of the KS scaling in this regime. The molecular gas in starbursts is likely not organized in units similar to galactic GMCs, and the basic physical properties and nature of this material are likely much more complex than those of local GMCs. Consequently, the nature of the \sigsfr - \siggas relation in this part of the KS diagram still awaits an explanation. Nonetheless, the considerations discussed in the previous section do suggest that even in these extreme conditions the role dense gas may still be the critical factor in controlling the SFR. 

However, a discrepancy remains between the local cloud and extragalactic observations. Namely, the gas depletion times ($\sim$ 220 - 250 Myrs) derived for local GMCs (and GMCs in NGC 300) differ significantly from those derived on 1 kpc scales of local disk galaxies (2-3 Gyr, e.g., Biegiel et al 2008). The depletion times derived for local GMCs are likely robust suggesting that either: 1) the extragalactic SFR calibrations underestimate the actual SFRs or 2) in addition to discrete GMCs, the extragalactic observations include diffuse molecular gas that is not participating in star formation or 3), some combination of these two effects. This issue clearly warrants more attention.

\section{Conclusions}
Comprehensive observations of nearby GMCs indicate that a Schmidt scaling relation of the form $\Sigma_{SFR} \propto \Sigma_{H_2}^n$, with n $\approx$ 2, exists {\it within} local molecular clouds. However, the observations also indicate that this relation alone does not provide a complete description of a cloud's star forming activity. Instead, the structure of a cloud is found to play a pivotal role in setting its global SFR. Moreover, observations clearly indicate that there is (and can be) {\it no} Schmidt relation connecting  \sigsfr and \siggas between GMCs, contrary to the case for galaxies. However, the integrated SFR is found to scale {\it linearly} with, and is most reliably traced by, the mass of {\it dense} gas in star forming GMCs. Moreover, this scaling relation appears to smoothly extend to global measurements of galaxies suggesting that the amount of dense gas in star forming regions sets the SFR in systems ranging from individual GMCs to entire galaxies. Deconstruction of the (resolved) KS law for local disk galaxies indicates that its functional form is largely a result of two factors: the unresolved observations of GMCs and the diluting effects of the inclusion of "inert" HI gas in measurements of \siggasp. Consequently, the KS relation provides little physical insight into the underlying process of star formation characterizing the molecular gas within galaxies. Nonetheless, indications are that the physics of star formation in distant galaxies and even through much of cosmic time may be similar in some fundamental respects to that we are witnessing today in local GMCs.

\section*{Acknowledgements}

\noindent  
I acknowledge the dedicated efforts of Joao Alves, Chris Faesi, Jan Forbrich, Marco Lombardi and Carlos Roman-Zuniga who made significant contributions to much of the research reviewed here.

\end{document}